\documentclass{jltp}

\usepackage{graphicx} % uncomment this line to include the graphicx package

\title{Stability and excitations of solitons \\ in 2D Bose-Einstein condensates}

\author{S. Tsuchiya$^1$, F. Dalfovo$^1$, C. Tozzo$^2$, and L. Pitaevskii$^{1,3}$}

\address{$^1$ CNR-INFM BEC Center and Physics Dept., University of Trento, Italy\\
$^2$ Scuola Normale Superiore, Piazza dei Cavalieri 7, I-56126 Pisa, Italy\\ 
$^3$ Kapitza Institute for Physical Problems, 119334 Moscow, Russia}

\runninghead{S. Tsuchiya, F.Dalfovo, C. Tozzo, and L. Pitaevskii}{Stability and 
excitations of solitons in 2D Bose-Einstein condensates}

\begin{document}

\maketitle

\begin{abstract}
The small oscillations of solitons in 2D Bose-Einstein condensates are investigated by 
solving the Kadomtsev-Petviashvili equation which is valid when the velocity of 
the soliton approaches the speed of sound. We show that the soliton is stable 
and that the lowest excited states obey the same dispersion law as the one of 
the stable branch of excitations of a 1D gray soliton in a 2D condensate. The role
of these states in thermodynamics is discussed. 
\end{abstract}

Solitons in Bose-Einstein condensates (BECs) have recently attracted much 
attention and have been studied both theoretically and experimentally. The 
Gross-Pitaevskii (GP) equation admits solitonic solutions corresponding to a 
local density depletion (i.e., gray and dark solitons). They have been created 
in 3D BECs\cite{Burger} and their decay into vortex rings has been 
observed\cite{Anderson}. In 2D, an interesting type of soliton corresponds 
to a self-propelled vortex-antivortex pair that, when it moves at a velocity 
close to the Bogoliubov sound speed, $c$, takes the form of a rarefaction 
pulse\cite{Jones}.  The energy $E$ and momentum ${\bf P}$ were calculated in 
Ref.\cite{Jones}, and it was found that the curve $E(P)$ is below the 
Bogoliubov sound line, approaching it in the low momentum limit. We investigate 
the excitation spectrum. Here 
we present our first results for solitons moving with velocity close to $c$, for 
which one can reduce the GP equation to the simpler Kadomtsev-Petviashvili (KP) 
equation\cite{Jones,KP}.  

Let us consider a 2D condensate with a soliton moving 
at velocity $V$ in the $x$-direction. If the 
density at large distances is $n_{\infty}$, one can define the healing length 
$\xi=\hbar/(2mgn_\infty)^{1/2}$, where $g$ is the mean-field coupling constant
and $m$ is the mass of the bosons. One can introduce the dimensionless 
variables $x \to \xi x$, $y \to \xi y$ and $t \to mt \xi^2/\hbar$, the normalized
order parameter $\Psi \to \sqrt{n_\infty} \Psi$ and the velocity $U=m\xi V/\hbar=
V/(c\sqrt{2})$ where $c=\hbar/(\sqrt{2}m\xi)$ is the sound speed. In the frame
moving with the soliton, the GP equation becomes\cite{Jones} 
%%%%%%%%%%%%%
\begin{eqnarray}
2i \frac{\partial}{\partial t}\Psi = 
-\nabla^2 \Psi + 2iU \frac{\partial}{\partial x}\Psi + 
\left(|\Psi|^2-1\right)\Psi \; . 
\label{GP}
\end{eqnarray}
%%%%%%%%%%%%%
The order parameter can be written as $\Psi=n^{1/2}e^{iS}$. In the 
limit of $V\to c$, one can introduce the small parameter 
$\varepsilon\equiv\sqrt{1-2U^2}$, so that $U \simeq 1/\sqrt{2} - 1/(2\sqrt{2}) 
\varepsilon^2$. The density depletion associated with the soliton becomes shallow 
and one can expand the density and the phase as
 %%%%%%%%%%%%%
 \begin{eqnarray}
 n&=&1 - \varepsilon^2f+\dots  \label{density}\\
 S&=&\varepsilon s +\dots\; \; .
 \end{eqnarray}
%%%%%%%%%%%%% 
To the lowest order in $\varepsilon$, the GP equation gives  
$\partial_x s=- \varepsilon f/\sqrt{2}$ and the function $f$ obeys
the KP equation 
%%%%%%%%%%%%%
\begin{eqnarray}
\frac{\partial}{\partial \tilde{x}}\left(\frac{\partial f}{\partial \tilde{t}}
+6f\frac{\partial f}{\partial \tilde{x}}+\frac{\partial^3 f}{\partial \tilde{x}^3}
\right) = \frac{\partial^2 f}{\partial  \tilde{y}^2} \; \; ,
\label{KP}
\end{eqnarray}
%%%%%%%%%%%%%%
where we have introduced the stretched variables $\tilde x= -\varepsilon x +
\varepsilon^3t/(2\sqrt{2})$, $\tilde y= \varepsilon^2 y/\sqrt{2}$ 
and $\tilde t=\varepsilon^3 t/(4\sqrt{2})$. 

\bigskip

{\it i) 1D gray soliton in 2D condensate.}

By assuming the stationary solution of (\ref{KP}) to be independent of 
$\tilde{y}$, Eq.~(\ref{KP}) reduces to the Korteweg-de Vries (KdV) 
equation and the solution is $f_0(\tilde{x}-2\tilde{t})=
{\rm sech}^2 [(\tilde{x}-2\tilde{t})/\sqrt{2}]$, corresponding to 
a gray soliton. The linear stability of 
this solution can be studied by taking $f(\tilde{x},\tilde{y},\tilde{t})=
f_0(\tilde{x}-2\tilde{t})+\psi(\tilde{x}-2\tilde{t})e^{i(\tilde{k}\tilde{y}
-\tilde{\omega} \tilde{t})}$ and linearizing Eq.~(\ref{KP}) in $\psi(\tilde{x})$. 
One gets 
%%%%%%%%%%%%%%
\begin{eqnarray}
\frac{d^4 \psi}{d \tilde{x}^4}+6\frac{d^2}{d\tilde{x}^2}\left( f_0 \psi \right)
-2\frac{d^2 \psi}{d \tilde{x}^2}+\tilde{k}^2\psi
=i\tilde{\omega}\frac{d\psi}{d\tilde{x}} \; .
\label{linKP1D}
\end{eqnarray}
%%%%%%%%%%%%%%
This equation was already solved by Zakharov\cite{Zakharov}, by applying the inverse 
scattering method, and also by Alexander {\it et al.}\cite{Alexander}. 
The excitations which are localized along $\tilde{x}$ and periodic along 
$\tilde{y}$ have dispersion law $\tilde{\omega}^{2} = 
(16 \tilde{k}^2/3\sqrt{3}) ( \tilde{k} - \sqrt{3}/2 )$.
For a given $\varepsilon$, the eigenfrequencies $\omega$ in the original 
units of the GP equation (\ref{GP}) are given by $\omega =\varepsilon^3 
\tilde{\omega}/(4\sqrt{2})$, and the wavevector transforms as 
$k=\varepsilon^2\tilde{k}/\sqrt{2}$, so that 
%%%%%%%%%%%%%%
\begin{eqnarray}
\omega^2=\frac{k^2}{6} \left(\sqrt{\frac{8}{3}}k-\varepsilon^2  \right) \; .
\label{gray}
\end{eqnarray}
%%%%%%%%%%%%%%
For $k > \varepsilon^2 \sqrt{3/8}$ the frequency is real, 
while for $k  < \varepsilon^2 \sqrt{3/8}$ it is imaginary, causing 
the instability of the soliton against bending.\cite{KP} 
In the limit $\varepsilon \to 0$, the region of instability becomes
vanishingly small and, for $k \gg \varepsilon^2$,  the stable branch of 
excitations becomes $\omega= (2/27)^{1/4}k^{3/2}$. It is worth noticing that 
this dispersion exhibits the same power law as for capillary waves on the
surface of a liquid.  

%%%%%%%%%%%%%%
\begin{figure}
\centerline{\includegraphics[width=10cm]{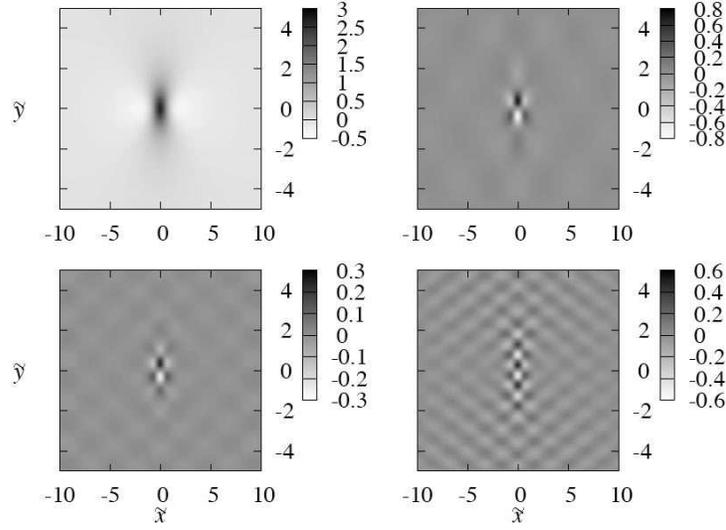}}
%\includegraphics{qfsfig1.ps}
%\framebox[1in]{\rule[1.125in]{0in}{1.125in}}
%\makebox[1in]{\rule[1.125in]{0in}{1.125in}}
\caption{Profile of the 2D soliton, Eq. (\protect\ref{2Dsoliton}), 
(upper left) and three examples of excited states with
$\tilde{\omega}=5.0$ (upper right), $\tilde{\omega}=11.8$ (lower left), and
$\tilde{\omega}=24.4$ (lower right). For the excitations, the real part of the
odd eigenfunctions is shown.}
\label{fig1}
\end{figure}
%%%%%%%%%%%%%%

\bigskip

{\it ii) 2D soliton.}

The solution of Eq.~(\ref{KP}) corresponding to a 2D soliton is\cite{Manakov}
%%%%%%%%%%%%%%
\begin{eqnarray}
f_0(\tilde{x}-2\tilde{t},\tilde{y}) = 4\frac{ \frac{3}{2}+2\tilde{y}^2-
(\tilde{x}-2\tilde{t})^2}{[\frac{3}{2}+2\tilde{y}^2
+(\tilde{x}-2\tilde{t})^2 ]^2} \; \; .
\label{2Dsoliton}
\end{eqnarray}
%%%%%%%%%%%%%%
This function is plotted in the upper left frame of Fig.~1. In the units of
the GP equation (\ref{GP}), the density profile of the soliton is found by 
inserting (\ref{2Dsoliton}) into (\ref{density}). Its size along $x$ and 
$y$ is of the order of $1/\varepsilon$ and $1/\varepsilon^{2}$, respectively.  
To the first order in $\varepsilon$, the momentum $P$ and the energy $E$ 
per unit length are\cite{Jones} 
%%%%%%%%%%%%%%
\begin{eqnarray}
P= \frac{8\pi}{3} \sqrt{2}\hbar n_{\infty} \xi\varepsilon \; \; , \; \; 
E= \frac{8\pi}{3m} \hbar^2n_{\infty}\varepsilon = cP \; .
\label{EP}
\end{eqnarray}
%%%%%%%%%%%%%%
As it must be, the soliton has a sound-like dispersion in the limit where 
KP equation is valid, that is, when $P \to 0$. 

Now we consider fluctuations of 
the form $f(\tilde{x},\tilde{y},t)=f_0(\tilde{x}-2\tilde{t},\tilde{y}) + 
\psi(\tilde{x} -2\tilde{t},\tilde{y})e^{-i\tilde{\omega} \tilde{t}}$ and 
linearize Eq.(\ref{KP}) by $\psi$. Thus, $\psi(\tilde{x},\tilde{y})$ 
satisfies
%%%%%%%%%%%%%%
\begin{eqnarray}
\frac{\partial^4 \psi}{\partial \tilde{x}^4}+6\frac{\partial^2}{\partial \tilde{x}^2}
\left( f_0\psi\right)-2\frac{\partial^2 \psi}{\partial \tilde{x}^2}
-\frac{\partial^2 \psi}{\partial \tilde{y}^2}=i \tilde{\omega} 
\frac{\partial \psi}{\partial \tilde{x}} \; \; .
\label{linKP2D}
\end{eqnarray}
%%%%%%%%%%%%%%
Looking for bound excited states (i.e., $\psi\to 0$ at large distances), it is 
convenient to integrate Eq.~(\ref{linKP2D}) twice along $\tilde{x}$ and introduce the
function $\phi(\tilde{x},\tilde{y})$ such that $\psi=\frac{\partial ^2 \phi}{\partial 
\tilde{x}^2}$.
The equation for $\phi$ is 
%%%%%%%%%%%%%%
\begin{eqnarray}
\frac{\partial^4 \phi}{\partial \tilde{x}^4} 
+ 6f_0\frac{\partial^2 \phi}{\partial \tilde{x}^2}
-2\frac{\partial^2 \phi}{\partial \tilde{x}^2}
-\frac{\partial^2 \phi}{\partial \tilde{y}^2}
=i\tilde{\omega} \frac{\partial \phi}{\partial \tilde{x}} \; \;.
\label{linKP2d2}
\end{eqnarray}
%%%%%%%%%%%%%%
We numerically solve this equation in a square box of size $L$, 
by expanding $\phi$ as
$\phi(\tilde{x},\tilde{y}) = \sum_{\nu,\mu} \phi_{\nu \mu} \chi_{1,\nu}(\tilde{x})
\chi_{2,\mu}(\tilde{y})$ with
$\chi_{1,\nu}(\tilde{x}) = L^{-1/2} e^{i2\pi \nu \tilde{x}/L}$ for  $|\nu| \le l$.
Concerning $\chi_{2,\mu}(\tilde{y})$ we note that Eq.~(\ref{linKP2D}) is invariant for 
$\tilde{y}\to -\tilde{y}$, so that the function $\phi$ is either even  or odd function 
of $\tilde{y}$. If it is even, then one can take $\chi_{2,\mu}(\tilde{y})= (2/L)^{1/2} 
\cos (2\pi \mu \tilde{y}/L)$ for $1 \le \mu \le l$ 
and $\chi_{2,0}(\tilde{y})=(1/L)^{1/2}$. 
If it is odd, one can take $\chi_{2,\mu}(\tilde{y})= (2/L)^{1/2} 
\sin (2\pi \mu \tilde{y}/L)$ for $1 \le \mu \le l$. One thus obtain the following
matrix equation 
%%%%%%%%%%%%%%
\begin{equation}
\left(-q_x^3-2q_x-q_y^2/q_x \right)\phi_{\nu \mu}+6\sum_{\nu^\prime, \mu^\prime}
q_x^\prime M_{\nu \mu , \nu^\prime \mu^\prime} \phi_{\nu^\prime \mu^\prime} =
\tilde{\omega} \phi_{\nu \mu} \; \; , 
\label{mequation}
\end{equation}
%%%%%%%%%%%%%%
with $q_x=2\pi \nu/L$, $q_y= 2\pi \mu/L$ and 
%%%%%%%%%%%%%%
\begin{equation}
M_{\nu \mu , \nu^\prime \mu^\prime} = \int_{-L/2}^{L/2} d\tilde{x} d\tilde{y} \ 
\chi_{1,\nu}^\ast\chi_{2,\mu}^\ast f_0 \chi_{1,\nu^\prime}\chi_{2, \mu^\prime} \; .
\end{equation}
%%%%%%%%%%%%%%
The size of the matrix is $N\times N$, where $N=2l(l+1)$ for $\phi$ even, and $N=2l^2$ 
for $\phi$ odd. Typical values in our calculations are $l=70$ and $L=60$. 
%%%%%%%%%%%%%%
\begin{figure}
\centerline{\includegraphics[width=8cm]{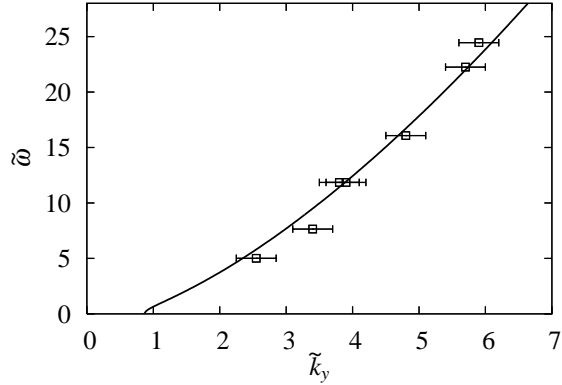}}
%\framebox[1in]{\rule[1.125in]{0in}{1.125in}}
%\makebox[1in]{\rule[1.125in]{0in}{1.125in}}
\caption{Excitation spectrum of the 2D soliton. Points are 
obtained from Eq. (\ref{linKP2d2}). 
The solid line is the dispersion law of the stable branch of 
excitations of a 1D gray soliton, Eq. (\ref{gray}).}  
\label{fig2}
\end{figure}
%%%%%%%%%%%%%%

It is found that all the eigenvalues are real. This is consistent with 
the results of Refs.\cite{Berloff,Kuznetsov} and shows the linear stability 
of the 2D soliton. One can expect this result also from pure kinematic 
considerations. An excited state is unstable if it can decay into several 
phonons.  Energy and momentum must be conserved in this process. It was 
proved by Iordanskii and one of the authors\cite{Iordanskii} that for a 
Bogoliubov-like spectrum the conservation laws can be satisfied only if 
the dispersion law $E(P)$ is above the sound line. Since the dispersion law 
of the 2D soliton is below the sound line, the soliton is stable.

We find stable localized modes with positive eigenvalues. Examples are 
shown in Fig.~1 for $\tilde{\omega}=5.0, 11.8$ and $24.4$. The function 
$\psi$ oscillates along $\tilde{x}$ and $\tilde{y}$, with pronounced maxima 
in the soliton region. Since the density profile of the soliton decays 
algebraically, $\psi(\tilde{x},\tilde{y})$ decays rather slowly. It is 
found that the real and imaginary part of $\psi$ are either even or 
odd functions of $\tilde{x}$ and each eigenvalue is degenerate, 
corresponding to even and odd functions of $\tilde{y}$. 

By looking at the oscillations of the eigenvectors $\psi$ in 
the soliton region, one can estimate the transverse and axial 
wavevectors, $\tilde{k}_x$ and $\tilde{k}_y$ respectively. In Fig.~2 we 
plot the calculated eigenfrequencies $\tilde{\omega}$ as a function of 
$\tilde{k}_y$ for the lowest bound states (points with error bars, where 
the error bars are of the order of the inverse of the axial size of the 
soliton). In the same figure we plot the dispersion law (\ref{gray}) 
of the stable branch of excitations of a 1D gray soliton. The spectrum 
is very similar. The 2D soliton have a discrete spectrum of bound 
states, due to its finite size, and all wavevectors are above the 
threshold for the instability of the 1D soliton.  

Notice that the soliton oscillations can be subject to damping 
by emission of sound waves. This means that the energy levels of Fig.~2 
have to be considered as resonance states.\cite{note}
However, there are reasons to believe that this damping is small and 
can be neglected both in the calculation of the energy levels and in 
the thermodynamic considerations below. We plan to investigate this 
damping in the future.

As it was noticed in Ref.\cite{Jones}, this 2D soliton can contribute to the 
specific heat $C$ of the 2D Bose gas. Indeed, if one applied the usual Bose 
statistics to the soliton branch, due to its sound-like dispersion (\ref{EP}),
its contribution would be $C\propto T^{2}$ as for phonons. The presence of 
the excited states of the soliton can change the situation. The inclusion of these
excitations is an open problem; here we just want to provide an argument 
which illustrates its nontriviality.

One can take into account the excitations of solitons by assuming that 
the solitonic branch of the spectrum depends on two quantum numbers: 
$\mathbf{P}$ and $k$. For example, within the limits of applicability 
of the dispersion law (\ref{gray}), one can write
\begin{equation}
E\left( \mathbf{P},k\right) =c\left| \mathbf{P}\right| +\beta k^{3/2} \; .
\end{equation}
One has to accept the hypothesis that the number of states in the 
interval $d^{2}P$ is, as
usual, $Sd^{2}P/\left( 2\pi \hbar \right) ^{2},$ where $S$ \ is the sample
area. On the contrary, the quantization rule for $k$ is fixed by the
length of the soliton in the $y$ direction, $R_y$, which behaves as $R_y\propto\xi /\varepsilon ^{2}$. Correspondingly, the number of states in the interval $dk$ is $\propto R_y dk/(2\pi)^3\propto (\xi /\varepsilon ^{2})dk$.
The energy of the gas is thus $\mathcal{E} \propto \int d^{2}Pdk\  
\varepsilon^{-2} E (\mathbf{P},k) [\exp(E(\mathbf{P},k)/T)-1]^{-1}$. A 
simple calculation shows that the relevant values are $E\propto \varepsilon 
\propto T,$ $P_{i} \propto T$ with $i=x,y$, and $k\propto T^{2/3}$. This
gives the specific heat $C\propto T^{2/3}$, which exceeds the phonon
contribution at low $T$. Of course, this result cannot be considered 
as a rigorous one, in particular because solitons occupy an area $\propto 1/\varepsilon^3$ and they can overlap at low temperature. Nevertheless it shows that the problem of the 
low-temperature specific heat of the 2D Bose gas deserves both 
theoretical and experimental investigation. A direct diagonalization 
of the Hamiltonian or Monte Carlo simulations at finite temperature 
could be suitable tools to test our semi-quantitative prediction. 

\bigskip

We thank C. Lobo, P. Pedri, N. Prokof'ev and D. Thouless for useful discussions.

\end{document}